\documentclass{superfri}
\usepackage{amsmath}
\usepackage{amssymb}
\usepackage{multirow}
\usepackage{subfigure}
\begin{document}
\author{Rio~Yokota\footnote{\label{kaust}King Abdullah University of Science and Technology (KAUST), Thuwal, 23955-6900, 
Saudi Arabia, \{rio.yokota,george.turkiyyah,david.keyes\}@kaust.edu.sa}, George~Turkiyyah\footnoteref{kaust}, 
David~Keyes\footnoteref{kaust}}
\title{Communication Complexity of the Fast Multipole Method and its Algebraic Variants}
\maketitle{}

\begin{abstract}
A combination of hierarchical tree-like data structures and data access patterns from fast multipole methods and hierarchical low-rank approximation of linear operators from $\mathcal{H}$-matrix methods appears to form an algorithmic path forward for efficient implementation of many linear algebraic operations of scientific computing at the exascale. The combination provides asymptotically optimal computational and communication complexity and applicability to large classes of operators that commonly arise in scientific computing applications. A convergence of the mathematical theories of the fast multipole and $\mathcal{H}$-matrix methods has been underway for over a decade. We recap this mathematical unification and describe implementation aspects of a hybrid of these two compelling hierarchical algorithms on hierarchical distributed-shared memory architectures, which are likely to be the first to reach the exascale. We present a new communication complexity estimate for fast multipole methods on such architectures.  We also show how the data structures and access patterns of $\mathcal{H}$-matrices for low-rank operators map onto those of fast multipole, leading to an algebraically generalized form of fast multipole that compromises none of its architecturally ideal properties.

\end{abstract}

\section{Introduction}
\subsection{Exascale features of the fast multipole method}
Wherever it can be applied, the fast multipole method (FMM) \cite{Greengard1987} has many appealing properties for extreme computing, beginning with its optimal computational complexity of $\mathcal{O}(N)$ for resolving the interdependence of $N$ degrees of freedom within a specifiable error tolerance. Its computationally expensive phases have extremely high flop/s to byte/s ratios, a data locality property known as ``computational intensity''  \cite{Williams2009}, up to two orders of magnitude better than the conventional sparse-matrix vector multiply kernel of Krylov and other purely algebraic solvers that do not take into account the low-rank mathematical structure of the operators being approximated. Meanwhile, the dominant kernels of FMM share with the matrix-vector multiply a level of computational concurrency that scales with problem size $N$. Furthermore, no all-to-all communication is required in an optimal implementation of FMM in a distributed memory environment of $P$ processes, and the $\mathcal{O}(\log P)$ messages exchanged are permitted, among themselves, a high degree of concurrency and asynchronicity. In short, fast multipole appears to be an ideal algorithm for massively distributed memory architectures with many cores sharing the local memory of a node -- the dominant design for contemporary extreme scientific computing.

The low complexity of FMM is accomplished by hierarchically clustering successively distant interactions to reduce the intrinsic $\mathcal{O}(N^2)$ arithmetic and $\mathcal{O}(P^2)$ communication complexity of an explicit interaction loop in a weak scaling implementation with $N/P$ degrees of freedom per process. A number of open-source libraries for fast multipole methods for distributed memory have been released \footnote{http://www.scafacos.de} 
\footnote{http://www.mrl.nyu.edu/\~{ }harper/kifmm3d/documentation/download.html} \footnote{https://bitbucket.org/rioyokota/exafmm-dev}. Hierarchical $N$-body methods have been at the core of many Gordon Bell Prizes \cite{Warren1992,Warren1997,Warren1998,Kawai1999,Hamada2009,Hamada2010,Rahimian2010,Ishiyama2012}. Looking inward in strong scaling within a fixed memory rather than outward in weak scaling with expanding memory, they have also been effectively implemented on GPGPUs \cite{Gumerov2008,Stock2008,Lashuk2009,Yokota2009,Gaburov2010a,Jetley2010,Burtscher2011,Yokota2011c,Bedorf2012,Langston2013}.  However, the mathematical theory of FMM is based upon the structure of the underlying operators, be they Laplace, Helmholtz, Stokes, elasticity, etc., and is based on forming and translating expansions of the Green's function, or resolvent operator.  One means to exploit the power of FMM for operators for which we do not possess Green's functions, but which are ``nearby'' in a spectrally equivalent sense, is to employ FMM (complemented with boundary elements as necessary to enforce finite domain boundary conditions) as a preconditioner inside a Krylov framework \cite{Yokota2014}. 

$\mathcal{H}$-matrix \cite{hackbusch99} theory is an alternative pathway for exploiting the increasingly low-rank structure of successively distant interactions in the context of the commonly arising constant-coefficient operators of scientific computing, and also for many others for which an explicit resolvent operator cannot be written down. There exist some quality open-source libraries for $\mathcal{H}$-matrices \footnote{http://bebendorf.ins.uni-bonn.de/AHMED.html} \footnote{http://www.hlibpro.com} \footnote{https://bitbucket.org/poulson/dmhm}, but they are essentially toolboxes for algorithmic experimentation. None yet approach or are essentially in their construction motivated by the extreme architectural requirements of the exascale.

Many adaptations of current workhorse algorithms, or fresh innovations, are required if currently anticipated exascale hardware is to be used near its potential in scientific computing, since our existing code base has been engineered primarily to squeeze out as many floating point operations as possible. Instead, algorithms must now focus on squeezing out synchronizations, memory storage, and memory transfers, while extra flops on locally cacheable data represent small costs in time and energy.

Today's scalable solvers, in particular, exploit convenient global synchronizations, for which top systems provide special hardware. The all-to-all exchange of scalar data on each node that all processes wait to complete, such as an inner product of globally distributed vectors, is an endangered idiom in architectures consisting of a billion threads. Even if the work imbalance between such synchronizing steps can be bounded, which is an incompletely solved challenge in distributed adaptive computations, the processors of the future cannot be expected to be performance-reliable. For example, dynamic clocking to maintain safe levels of heat generation or dynamic correction of errors due to low signal-to-noise ratios in energy efficient hardware, will cause completion times for equal work to vary in unpredictable ways among identical processors. After decades of algorithm refinement during a period of programming model stability with bulk synchronous processing (BSP) \cite{Valiant1990}, new programming models and new algorithmic capabilities (to take full advantage of the potential of exascale simulation in such areas as data assimilation, inverse problems, and uncertainty quantification) must be co-designed with the hardware.

There are numerous constraints on exascale algorithms, chief among which are minimizing memory storage and minimizing frequency of deep memory access. Roadmaps for exascale (e.g., \cite{Kogge2008}) show that energy requirements of memory accesses at increasing distance or depth will grow by orders of magnitude relative to the cost of performing a floating point operation between the operands. Further, storage will dominate the cost of acquisition of exascale hardware. Hence, algorithmic solutions are directly constrained by operating and acquisition costs to be parsimonious in memory use.

For the Laplace and Laplace-like operators that commonly arise in elliptic, time-implicit parabolic, and frequency-domain hyperbolic PDEs, algebraic multigrid (AMG) scales efficiently in proportion to available memory on hundreds of thousands of rigidly schedulable, tightly coupled cores for systems with billions of unknowns \cite{Baker2012}. However, for constant coefficient problems the fast multipole method (FMM) is asymptotically superior in complexity and tolerates less synchronization. Algebraic fast multipole (AFM) based on hierarchical matrix decompositions may extend the performance robustness of FMM to the coefficient variability of AMG and improve upon memory requirements, and therefore exemplifies the adaptations that many algorithms must make.

In Section 1.2, we recall a hierarchical feature of contemporary architecture and accompanying constraints that motivate wider exploitation of FMM. In section 1.3, we recall the advantages of several hierarchical algorithms and we motivate interest in an algebraic variant of fast multipole. Section 2 presents a new result on FMM communication complexity with respect to three types of distributions that are illustrative of applications in 3D. Section 3 describes the $\mathcal{H}^2$ formulation of $\mathcal{H}$-matrices and highlights how its data structures and communication patterns map onto those of the FMM and how the matrix decomposition is formed. In Section 4, we conclude this work in progress with a consideration of open questions.

\subsection{Architectural implications for exascale algorithms}
Due to the leveling of clock speeds of CMOS processors primarily for reasons of energy consumption and concomitant heat dissipation, the first CMOS-based exascale machines will consist of approximately one billion threads executing at about 1 GigaHertz. They will likely be configured approximately as one million nodes of one thousand cores each: a GigaHertz-MegaNode-KiloCore machine. Perhaps the number of cores per fixed memory node will not exceed one hundred, in which case ten million nodes will be required. Weak scaling has proved to be an easier paradigm than strong scaling to meet efficiently because extra memory and extra memory bandwidth come proportionally with the increase in cores. The high cost of this memory is one of the main pressures in favor of manycore and GPGPU architectures. Forecasts of the architecture of exascale may be found in \cite{Kogge2008} and \cite{Dongarra2011}. Under any form of scaling, algorithms are favored that reduce the per flop requirements of storage and memory bandwidth, and increase the uniformity and predictability of the flops that are ultimately executed.

Bad and good news simultaneously lurk behind every aspect of the evolution of exascale architecture. Programmers will have to explicitly control more of the data motion, since it carries the highest energy cost in the computational environment.  However, they will be given tools that some have craved to better control the vertical (replication-oriented) data motion. The horizontal (interprocessor) data motion has been successfully under programmer control in the form of message passing for over 25 years. Vertical replication has generally been handled beneath the level of programmer concern. However, recently with the rolling of data in and out of accelerators, programmers have embraced this control, even if they bemoan the slow transmission speed of today's channels. Authors of performance-seeking tree-based codes are already in the habit of managing their data in great detail, to the extent that dynamic runtime systems may have little to improve upon \cite{AbdulJabbar2014}.

Today's optimal algorithms have evolved, as mentioned above, to squeeze out flops, whereas the premium at the exascale is to squeeze out memory storage and accesses. However, whereas storage-optimal methods may not be computationally optimal, computationally optimal methods tend to be storage-optimal because the amount of relevant data cannot be greater than that touched in executing the flops. Therefore, computationally optimal algorithms are the first places to look for memory parsimonious exascale kernels.

Other drivers of exascale algorithmic innovation, generally, include exploitation of adaptive precision, as a means of reducing storage, and algorithmic-based fault tolerance, including detecting and handling errors within the user application, rather than making fault tolerance a hardware responsibility. In some sense, optimal fast multipole and $\mathcal{H}$-matrix formulations control accuracy in a way that reduces the opportunity and the pressure for adaptive precision -- very few bytes are wasted by mathematical overresolution compared to other formulations of the same underlying continuous mathematics.  As for algorithmic-based fault tolerance (ABFT), FMM and $\mathcal{H}$-matrices may lack an advantage of multigrid and Krylov solvers, with their fresh periodic computations of residuals to drive corrections in the latter. However, ABFT for FMM and $\mathcal{H}$-matrices is in its infancy. Today's hardware can still be treated as reliable, and among today's optimal hierarchical methods, FMM and $\mathcal{H}$-matrices excel in concurrency, tolerance of asynchronicity, arithmetic intensity, and the ability to tune SIMT data sizes to natural boundaries in the hardware hierarchy.

\subsection{Hierarchical algorithms}
A diverse collection of algorithms with optimal arithmetic complexity -- linear or at most log-linear in the number of degrees of freedom -- share the key characteristic of being hierarchical.  We list them here in their simplest forms; each one is, of course, the subject of decades of research that fill books.

The fast Fourier transform (FFT) replaces $\mathcal{O}(N^2)$ multiply-add operations with $\mathcal{O}(N \log N)$ with a constant as low as 5 for favorable radix. Geometric multigrid for solving the Poisson Dirichlet on a uniform 3D grid with second-order finite differences replaces $\mathcal{O}(N^{7/3})$ multiply-add operations for a Gaussian band solver on a naturally ordered version of the problem with $\mathcal{O}(N \log N)$ with a constant as low as about $6$ for solution to truncation error.  Sparse grids represent sufficiently smooth functions on a $d$-dimensional of resolution $N$ on a side, which would require $N^d$ to store, with $\mathcal{O}(N (\log N)^{d-1})$ storage at a cost in accuracy that is only logarithmically degraded. Like the FFT and multigrid, the ``combination form'' of sparse grids is able to work with data structures at each level of the hierarchy that have the same simple Cartesian structure of the original. All three algorithms therefore generate a logarithmic number of hierarchically coarsened versions of the original. (Multigrid may require a special solve for the coarsest grid.) 

Fast multipole reduces an $\mathcal{O}(N^2)$ summation to an $\mathcal{O}(N \log N)$ or $\mathcal{O}(N)$ complexity through a hierarchy of tree-based operations by distinguishing between near interactions that must be treated directly and recursively coarsened far interactions. Its complexity of implementation, even in serial, is therefore somewhat greater than the hierarchical methods above, but like the others, the principal feature is the recursive generation of a self-similar collection of problems. The $\mathcal{H}$-matrix format is suitable for a variety of operations, including matrix-vector multiplication, matrix inversion, matrix-matrix multiplication, matrix-matrix addition, etc., and has recently been generalized to tensors. In its classical $\mathcal{H}$ and $\mathcal{H}^2$ forms, the asymptotic complexity $\mathcal{O}(N^2)$ of a matrix-vector multiply is reduced to $\mathcal{O}(N \log N)$ or $\mathcal{O}(N)$, respectively. Fast Multipole Methods (FMM) and $\mathcal{H}$-matrices share many common features that arise from their hierarchical nature. The former is geometric while the latter is algebraic, but the two methods have similar work-flow and data structures. Many parallelization techniques that have been developed for the FMM can be applied to its algebraic variant. The matrix-free nature of FMM and the larger prefactor of its computational complexity are the sources of its strong arithmetic intensity.  Being matrix-free also reduces the amount of communication. 

Other methods deserve to be mentioned in this context even if they are not strictly optimal by the definition above. Nested dissection ordering and its graph partitioning generalization, when applied to algebraic systems generated from local PDE discretizations, create a recursive sequence of separators of decreasing size and leave behind independent blocks. Employing low-rank representations for the separator blocks preserves sparsity and lowers the overall factorization costs both in memory and flops. The dense Schur complements that arise recursively in many initially sparse discretizations are amenable to low-rank decompositions. For elliptic problems, large amounts of compression are possible and the resulting factorization can be used as a nearly optimal-order direct solver \cite{xia10}.

These hierarchical algorithms are vital to the optimal performance of many scientific codes, as kernels in simulations based on formulations of partial differential equations, integral equations, and interacting particles. Successfully migrating these kernels to the computational environment of the exascale will create a path that many full applications can follow, just as a generation earlier, dense and sparse linear algebra libraries led applications into efficient use of distributed memory and message passing.

\section{Communication Complexity of Fast Multipole Methods}
Communication becomes the bottleneck for any algorithm as it approaches the limit of its parallel scalability. Therefore, communication complexity is what distinguishes algorithms that scale from ones that do not. It is well known that FMM has $\mathcal{O}(N)$ \textit{arithmetic} complexity, but relatively little attention has been given to its \textit{communication} complexity. In this section we provide new upper bounds for the communication complexity of FMM, first for the uniform case. We then extend the complexity analysis to the nonuniform case and prove that the same upper bound holds. We define $N$ as the sum of the number of particles on all processes and $P$ as the number of processes. These are the only two variables in the current analysis of the communication complexity. We do not consider the communication during the initial partitioning phase in the present analysis.

\begin{table}[b]
\caption{Communication complexity of FMM.}
\label{tab:complexity}
\begin{center}
\begin{tabular}{|c|c|c|c|c|c|c|}
\hline
Reference & \multicolumn{2}{c|}{Processes} & \multicolumn{2}{c|}{Data per Process} & Communication complexity\\
\hline \hline
Teng \cite{Teng1998} & \multicolumn{2}{c|}{$\mathcal{O}(P)$} & \multicolumn{2}{c|}{$\mathcal{O}\left((N/P)^{2/3}(\log N+\mu)^{1/3}\right)$} & $\mathcal{O}\left(P(N/P)^{2/3}(\log N+\mu)^{1/3}\right)$\\
\hline
Lashuk \textit{et al.} \cite{Lashuk2009} & \multicolumn{2}{c|}{$\mathcal{O}(\sqrt{P})$} & \multicolumn{2}{c|}{$\mathcal{O}\left((N/P)^{2/3}\right)$} & $\mathcal{O}\left(\sqrt{P}(N/P)^{2/3}\right)$\\
\hline
\multirow{2}{*}{Ibeid \textit{et al.} \cite{Ibeid2014}} & Global & Local & Global & Local & Global + Local \\
\cline{2-6}
\multirow{2}{*}{} & $\mathcal{O}(\log P)$ & $\mathcal{O}(1)$ & $\mathcal{O}(1)$ & $\mathcal{O}\left((N/P)^{2/3}\right)$ & $\mathcal{O}\left(\log P+(N/P)^{2/3}\right)$ \\
\hline
\end{tabular}
\end{center}
\end{table}

\subsection{Uniform distribution} \label{sec:uniform}
Analysis of communication complexity of FMM has been performed previously by Teng \cite{Teng1998}, Lashuk \textit{et al.} \cite{Lashuk2009}, and Ibeid \textit{et al.} \cite{Ibeid2014}. The communication complexity of these different studies is shown in \tabref{tab:complexity}. ``Processes'' is the number of processes involved in the communication. Teng assumes a nonuniform distribution, while the other two focus on uniform distributions, but this is not what causes the difference in the complexity. We will show in section \ref{sec:nonuniform} that the tightest upper bound of Ibeid \textit{et al.} still holds for the nonuniform case. In this section we will first prove the upper bounds for the uniform case, while describing the reason for the discrepancies between the three. The three papers use different terminology so we provide \tabref{tab:terminology} to map the correspondence.

\begin{table}[b]
\caption{Correspondence of terminology.}
\label{tab:terminology}
\begin{center}
\begin{tabular}{|c|c|c|}
\hline
Teng \cite{Teng1998} & Lashuk \textit{et al.} \cite{Lashuk2009} & Ibeid \textit{et al.} \cite{Ibeid2014} \\
\hline
near-field graph & U-list & P2P-list \\
\hline
far-field graph & V-list & M2L-list \\
\hline
box & box & cell \\
\hline
\end{tabular}
\end{center}
\end{table}

There are two improvements between the communication complexity of Teng $\mathcal{O}\left(P(N/P)^{2/3}(\log N+\mu)^{1/3}\right)$ and Lashuk \textit{et al.} $\mathcal{O}\left(\sqrt{P}(N/P)^{2/3}\right)$. The first is $P$ changing to $\sqrt{P}$ and the second is the $(\log N+\mu)^{1/3}$ disappearing. The $P$ changing to $\sqrt{P}$ is due to $\displaystyle\sum_i^{\log P-1}2^i=\mathcal{O}(P)$ on the bottom of page 650 in Teng \cite{Teng1998} changing to $\displaystyle\sum_i^{\log P-1}min(2^{\log P-i-1},2^i)=\mathcal{O}(\sqrt{P})$ on the right side of page 7 in Lashuk \textit{et al.} \cite{Lashuk2009}. This improvement is made possible by the hypercube reduce and scatter scheme shown in Algorithm 3 in the same paper, where processes are paired in a $(\log P)$-dimensional hypercube. The $(\log N+\mu)^{1/3}$ factor in Teng stems from the proof of Lemma 4.8 where he assumes that there could be $\mathcal{O}(\log N+\mu)$ neighbors in the near-field graph if a highly refined leaf box existed next to a large leaf box. Such cases do not exist for a uniform distribution so it is easy to prove that this factor disappears in this case. Furthermore, we will show in section \ref{sec:nonuniform} that it is still possible to bound the neighbors in the near-field list to $\mathcal{O}(1)$ for a nonuniform distribution by using a 2:1 refinement constraint \cite{Sundar2008} during the tree construction.

\begin{figure}[t]
\hspace{-10mm}
\includegraphics[width=1.1\textwidth]{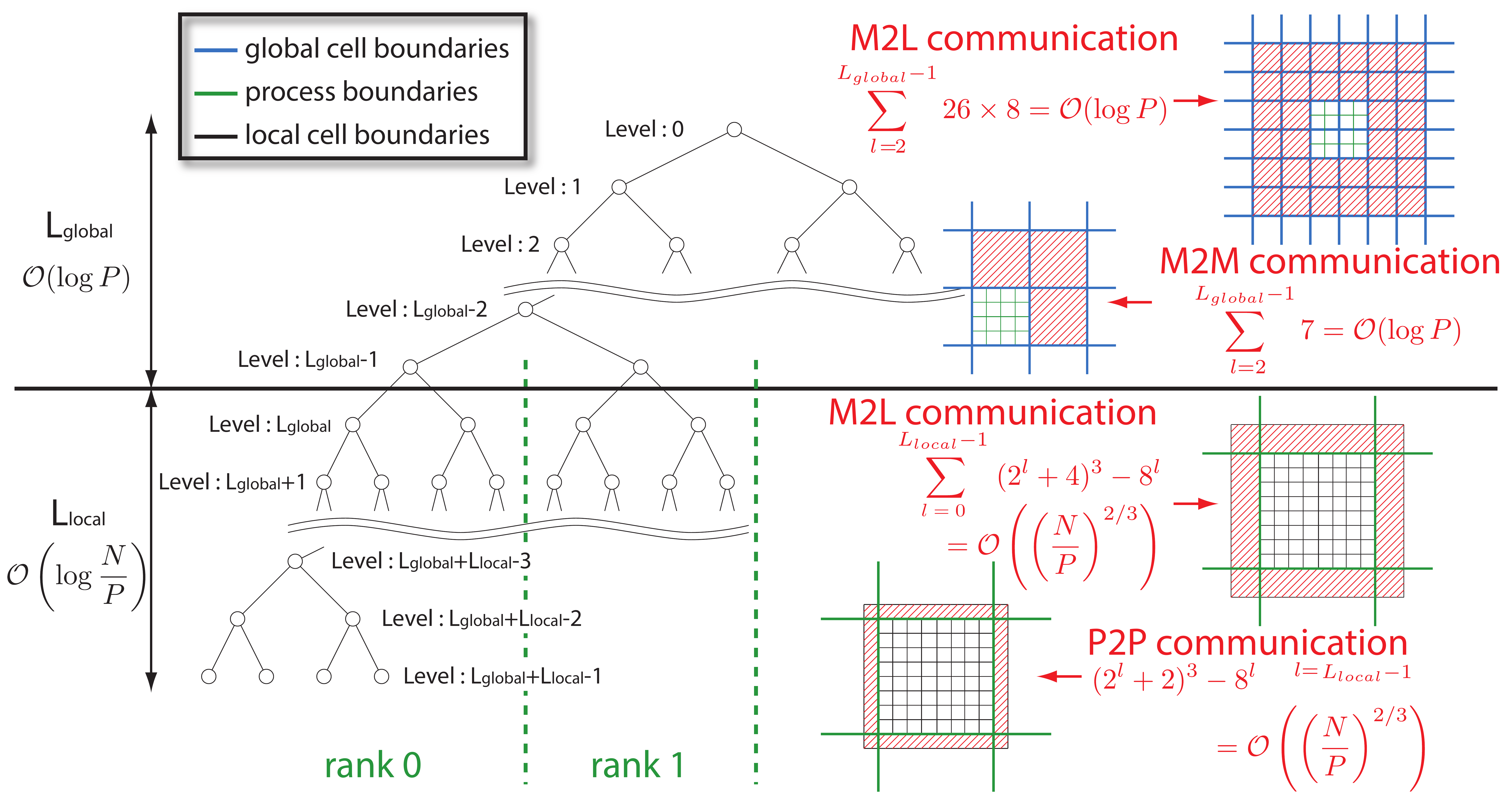}
\caption{Communication patterns and complexities of FMM.}
\label{fig:communication}
\end{figure}

We now focus on the two discrepancies between the communication complexity of Lashuk \textit{et al.} $\mathcal{O}\left(\sqrt{P}(N/P)^{2/3}\right)$ and Ibeid \textit{et al.} $\mathcal{O}\left(\log P+(N/P)^{2/3}\right)$. The first difference is $\sqrt{P}$ becoming $\log P$ and the second is the product changing to a sum. The change from a product to a sum comes from the separation of the global tree from the local tree as shown in \figref{fig:communication} and \tabref{tab:complexity}. The global tree is formed from the hierarchical grouping of processes, which has $P$ leaf nodes and a $\mathcal{O}(\log P)$ depth. The local tree is formed from local particles that belong to the process and has $\mathcal{O}(\log (N/P))$ depth. For a uniform distribution with a full tree and when the number of processes is a power of two, it is very easy to see that the leaf of the global tree is the root of the local tree as shown in \figref{fig:communication}. We will show in section \ref{sec:nonuniform} that this separation between the global tree and local tree is still possible for nonuniform distributions as well.

We separate the global tree from the local tree because they have different communication complexity as shown in \tabref{tab:breakdown} and also \figref{fig:communication}. ``Processes'' is the total number of processes to communicate with during a given phase. In the global tree, one octree cell is owned by many processes, whereas in the local tree many octree cells are owned by a single process. Therefore, the redundancy of the information of octree cells in the global tree increases exponentially as the level gets coarser. Exploiting this redundancy and getting rid of the all-to-all type communication pattern is what brings the $\sqrt{P}$ in Lashuk \textit{et al.} \cite{Lashuk2009} down to $\log P$ in Ibeid \textit{et al.} \cite{Ibeid2014}. The ability to remove the all-to-all type communication altogether may seem counterintuitive since information from every process \textit{is} required by every other process for the construction of the local essential tree. The key is to use the reduction in the M2M operation and the redundancy at the coarse levels to limit the number of processes that need to communicate directly. 

\begin{table}[b]
\caption{Breakdown of communication in Ibeid \textit{et al.} \cite{Ibeid2014}.}
\label{tab:breakdown}
\centering
\begin{tabular}{|l|c|c|c|c|}
\hline
 & Processes & Cells per level & Cells per Process & Communication\\
\hline \hline
Global M2L & $\displaystyle\sum_i^{\log P}26$ & $26\times8$ & $8$ & $\mathcal{O}(\log P)$\\
\hline
Global M2M & $\displaystyle\sum_i^{\log P}7$ & 7 & $1$ & $\mathcal{O}(\log P)$\\
\hline
Local M2L & $26$ & $(2^i+4)^3-8^i$ & $\displaystyle\sum_i^{\log_8(N/P)}(2^i+4)^3-8^i$ & $\mathcal{O}((N/P)^{2/3})$\\
\hline
Local P2P & $26$ & $(2^i+2)^3-8^i$ & $(2^{\log_8(N/P)}+2)^3-8^{\log_8(N/P)}$ & $\mathcal{O}((N/P)^{2/3})$\\
\hline
\end{tabular}
\end{table}

As an example, let us first consider the communication phase for the M2M operation in the global tree shown in \figref{fig:communication}. At the leaf nodes of the global tree one local root cell exists on each process, and the M2M operation will take each cell from the neighboring 8 processes and reduce this information to the parent node. This requires communication with the 7 neighboring processes. The tree structure shown in \figref{fig:communication} is a binary tree, which would be an octree in 3D. Similarly, the illustrations of the geometric partitions in \figref{fig:communication} are in 2-D but we are actually considering a 3-D octree. After the M2M operation at the leaf node (level $L_{global}$) of the global tree, the neighboring 8 processes have the same information about the node at level $L_{global}-1$. Therefore, when performing the M2M operation at the next level, one only needs to communicate with $7$ processes instead of $7\times 8$ due to the $8$-fold redundancy. If we apply this logic recursively, we see that the M2M communication will always require communication with $7$ processes. The $7$ processes to communicate with will move farther and farther away as we go up the tree, so we must sum up the number of processes at each level to get to total number of processes to communicate with. There are $\mathcal{O}(\log P)$ levels in the global tree and there is always one cell worth of data that is being sent, so the communication complexity of the global M2M phase is $\mathcal{O}(\log P)$ as shown in \tabref{tab:breakdown}.

Similar logic can be applied to the communication for the global M2L operations. Even though the number of octree cells contained in each process grows exponentially at coarser levels of the global tree, the redundancy of information also increases at the same exponential rate if we use the M2M communication described above. Therefore, the \textit{unique} information that must be communicated during the M2L phase remains constant at each level of the global tree. For the standard definition of neighbors, the M2L phase requires two cells worth of a halo region to be communicated. This results in $26\times 8$ cells to be communicated from $26\times 8$ neighboring processes. However, the 8 child cells that arrived during the M2M communication can be used to reduce the number of processes to communicate with to $26$, while each process now sends $8$ cell's worth of data as shown in \tabref{tab:breakdown}. Similar to the M2M phase, there are $\displaystyle\sum_i^{\log P}26$ processes to communicate with in total. Therefore, the communication complexity of the global M2L phase is $\mathcal{O}(\log P)$.

For the local M2L and P2P phases, it is only necessary to communicate with the same $26$ neighboring processes regardless of the level, so the number of processes to communicate with is $\mathcal{O}(1)$ instead of $\mathcal{O}(\log P)$. However, unlike the global tree communication, the local tree communication requires more cells to be sent as the level of the local tree gets finer. The M2L phase has two cells worth of a halo region and the P2P phase has one cell worth of a halo region that needs to be communicated, so the width of these halos are constant but their lengths are not. For the M2L halo, having two cells on each side adds four cells per dimension so we have $(2^i+4)^3-8^i$ cells to send at the $i$-th level. For the P2P halo, we have one cell on each side so the number is $(2^i+2)^3-8^i$. The halo size is basically the surface to volume ratio, which means that the complexity is $\mathcal{O}(N/P)^{2/3}$. It can also be seen from \tabref{tab:breakdown} that summing for powers of four up to a base eight logarithm gives $\displaystyle\sum_i^{\log_8(N/P)}4^i=(N/P)^{\log 4/\log 8}=(N/P)^{2/3}$.

Looking back at \tabref{tab:complexity}, we now see that it is the separation of the global tree from the local tree that turns the product in $\mathcal{O}(\sqrt{P}(N/P)^{2/3})$ of Lashuk \textit{et al.} to the sum in $\mathcal{O}(\log P+(N/P)^{2/3})$. In other words, there are a $\mathcal{O}(1)$ number of neighboring processes that require $\mathcal{O}(N/P)^{2/3}$ data during the local P2P and M2L communications. For the global communications the amount of data to send per process pair is $\mathcal{O}(1)$ but there are a constant number of processes to communicate with that are different at each level of the global tree, which has a depth of $\mathcal{O}(\log P)$. This is a significant improvement on the upper bound of the communication complexity of FMM with respect to Teng \cite{Teng1998} and Lashuk \textit{et al.} \cite{Lashuk2009}. We will extend this analysis for uniform distributions to nonuniform distributions in the following subsection.

\subsection{Nonuniform distribution} \label{sec:nonuniform}
\begin{figure}[h]
\centering
\subfigure[Uniform]{
\includegraphics[width=0.25\textwidth]{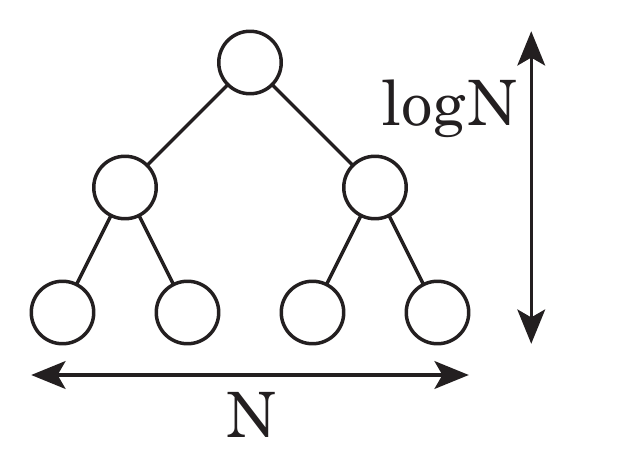}\label{fig:uniform}}
\hspace{10mm}
\subfigure[Nonuniform]{
\includegraphics[width=0.32\textwidth]{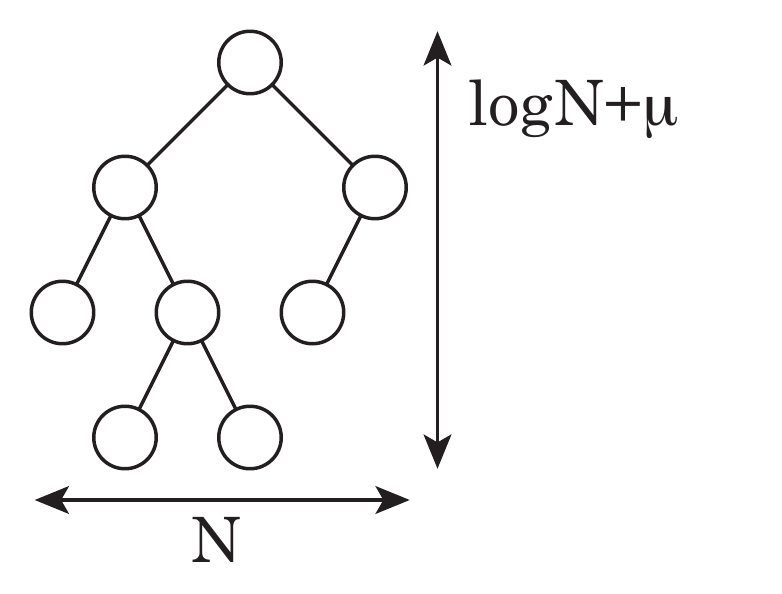}\label{fig:nonuniform}}
\hspace{2mm}
\subfigure[Pathological]{
\includegraphics[width=0.2\textwidth]{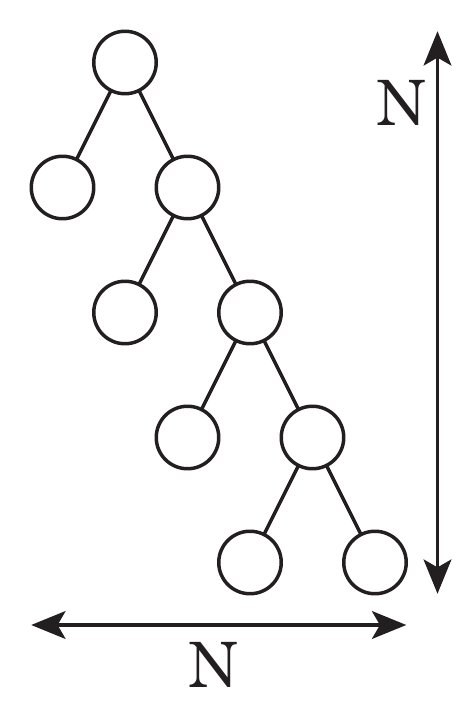}\label{fig:pathological}}
\caption{Refinement distributions reflected in tree data structures.}
\label{fig:distribution}
\end{figure}
The complexity analysis for the uniform distribution with a full octree can be extended to the nonuniform distribution with an adaptive tree. Before discussing the extension to nonuniform distributions, we first define the type of nonuniform distribution in which we are interested. For this purpose, three different types of tree structures are shown in \figref{fig:distribution}. The full tree that corresponds to uniform distribution as discussed in the previous subsection is shown in \figref{fig:uniform}. The adaptive tree that results from a $\mu$-nonuniform distribution \cite{Teng1998} is shown in \figref{fig:nonuniform}. We will discuss the $\mu$-nonuniform distribution in this subsection. The $\mu$-nonuniform distribution is in contrast to a pathological nonuniform distribution that results in a $\mathcal{O}(N)$ depth tree shown in \figref{fig:pathological}.

When considering nonuniform distributions, it is necessary to exclude pathological cases like the one in \figref{fig:pathological}.
Fortunately, such a distribution will not occur in practice since it corresponds to exponentially increasing spatial refinement throughout the entire domain. One could accidentally produce such a distribution by using a naive adaptive mesh refinement near a singularity, but this is obviously not a sound technique for the numerical integration of singular functions. Furthermore, for such pathological cases the FMM will reduce to a direct $N$-body method since the depth of the tree becomes $\mathcal{O}(N)$. At every level of the tree, all cells are direct neighbors of each other so the far-field approximation is not valid between any of them. Therefore, all pairs of particles will be calculated by direct summation in this case. Hence, the hypothesis for the arithmetic complexity of FMM being $\mathcal{O}(N)$ excludes such cases to begin with, so they will not be considered in the current communication complexity analysis.

\begin{figure}[h]
\centering
\includegraphics[width=0.8\textwidth]{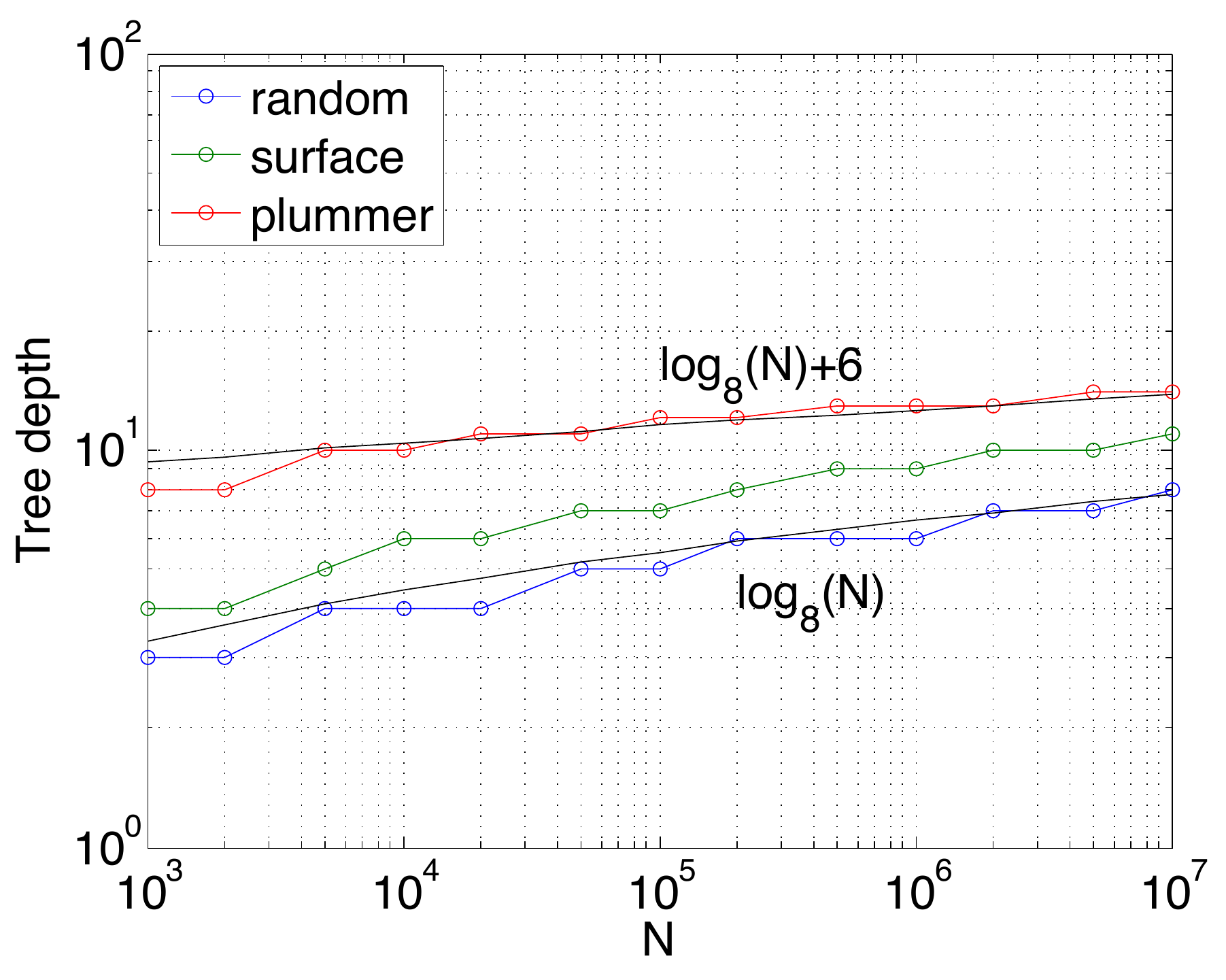}
\caption{Tree depth as a function of number of particles $N$ for different distributions. ``random'' is a random distribution of particles in a cube, ``surface'' has points only on the surface of a sphere, and ``Plummer'' \cite{Plummer1911} has high concentration of particles in the center of the domain. They all exhibit $\mathcal{O}(\log N)$ behavior but with a different constant. The maximum number of particles per leaf cell was set to $16$ for these plots.}
\label{fig:depth}
\end{figure}

The communication complexity for the uniform case is extended to the nonuniform case by adopting the definition of $\mu$-nonuniform distributions \cite{Teng1998}, which is depicted in \figref{fig:nonuniform}. For $\mu$-nonuniform distributions, the depth of the tree is still $\mathcal{O}(\log N)$ but has a constant number of additional levels that come from the nonuniformity. In \figref{fig:depth} we show the depth of the tree as a function of the number of particles $N$ for three types of distributions. ``random'' is a random distribution of particles in a cube, which is representative of the distribution of atoms in a molecular dynamics simulation. ``surface'' has points only on the surface of a sphere and is representative of a boundary integral calculation. ``Plummer'' has high concentration of particles in the center of the domain, which is common in cosmological simulations. The ``random'' distribution results in an almost full tree and is the most uniform among the three. The ``surface'' distribution has inter-particle spacing that grows as $\mathcal{O}(N^\frac{1}{d-1})$ for a $d$ dimensional simulation. Therefore, it still has $\mathcal{O}(\log N)$ depth but with a different constant. The ``Plummer'' distribution also has $\mathcal{O}(\log N)$ depth but with an even larger constant. FMM applications can be categorized into either of these three types of distributions, so we will assume the $\mu$-nonuniform distribution for the following analysis of communication complexity of FMM.

The maximum depth of the tree has further constraints that stem from the finite precision in numerical simulations. The first limit comes at 22 levels of an octree, where the 64-bit unsigned integer will overflow $2^{64} < 8^{22}$. It is possible to build deeper trees by using multiple integers to store the Morton/Hilbert keys. The next limit occurs at 53 levels, where the number of significand (mantissa) bits will not be enough to distinguish the two points expressed in double precision floating point numbers. Say for example, we have two particles with coordinates $\mathbf{x}_i$ and $\mathbf{x}_{i+1}$ with a distance $|\mathbf{x}_{i+1}-\mathbf{x}_i|/|\mathbf{x}_i|<1/2^{53}$. The double precision floating point value for the coordinates of these two particles will be identical because the difference will be in the 54th mantissa bit, which does not exist. Therefore, even if we use multiple 64-bit integers to store the Morton/Hilbert key, it would not be possible to build a tree structure with over 53 levels, because particles are indistinguishable past that level and cannot be correctly binned into to smaller sub-cells. Actually, the round-off error in the P2P kernel will become unacceptable far before we reach this level of refinement.

With this definition of $\mu$-nonuniform distributions, we revisit the differences between the communication complexity of Teng \cite{Teng1998}, Lashuk \textit{et al.} \cite{Lashuk2009}, and Ibeid \textit{et al.} \cite{Ibeid2014} shown in \tabref{tab:complexity}, and see if they are valid for the $\mu$-nonuniform case. The first difference between Teng and Lashuk \textit{et al.} is the change from $P$ to $\sqrt{P}$. There is no assumption of uniformity in the hypercube reduce and scatter communication that yields the $\sqrt{P}$ factor, so this communication scheme should be directly applicable to Teng's analysis, which will change $P$ to $\sqrt{P}$. We have mentioned in section \ref{sec:uniform} that the $(\log N+\mu)^{1/3}$ factor comes from the assumption that there could be $\mathcal{O}(\log N+\mu)$ neighbors in the near-field graph if a highly refined leaf box existed next to a large leaf box. The far-field graph does not contribute to the $\mathcal{O}(\log N+\mu)$ factor because the proof of Lemma 4.8 in Teng \cite{Teng1998} shows that all non-leaf-boxes have a in-degree bounded by a constant. Therefore, all we need to prove is that the near-field graph can be bounded by a constant for any $\mu$-nonuniform distribution in order to extend the communication complexity of Lashuk \textit{et al.} to a $\mu$-nonuniform case. The 2:1 balance refinement of octrees by Sundar \textit{et al.} \cite{Sundar2008} will yield precisely such a $\mathcal{O}(1)$ bound on the near-field graph. Therefore, the communication complexity $\mathcal{O}\left(\sqrt{P}(N/P)^{2/3}\right)$ of Lashuk \textit{et al.} is valid for $\mu$-nonuniform distributions if the hypercube reduce and scatter communication and 2:1 refinement are used.

\begin{table}[b]
\caption{Breakdown of communication for the $\mu$-nonuniform case.}
\label{tab:nonuniform}
\centering
\begin{tabular}{|l|c|c|c|c|}
\hline
 & Processes & Cells per level & Cells per Process & Communication\\
\hline \hline
Global M2L & $\displaystyle\sum_i^{\log P}\mathcal{O}(1)$ & $\mathcal{O}(1)$ & $\mathcal{O}(1)$ & $\mathcal{O}(\log P)$\\
\hline
Global M2M & $\displaystyle\sum_i^{\log P}\mathcal{O}(1)$ & $\mathcal{O}(1)$ & $\mathcal{O}(1)$ & $\mathcal{O}(\log P)$\\
\hline
Local M2L & $\mathcal{O}(1)$ & $\mathcal{O}(4^i)$ & $\displaystyle\sum_i^{\log_8(N/P)}\mathcal{O}(4^i)$ & $\mathcal{O}((N/P)^{2/3})$\\
\hline
Local P2P & $\mathcal{O}(1)$ & $\mathcal{O}(4^i)$ & $\mathcal{O}(4^{\log_8(N/P)})$ & $\mathcal{O}((N/P)^{2/3})$\\
\hline
\end{tabular}
\end{table}

The next task is to prove that the change from $\mathcal{O}\left(\sqrt{P}(N/P)^{2/3}\right)$ to $\mathcal{O}(\log P+(N/P)^{2/3})$ is valid for $\mu$-nonuniform distributions. For a $\mu$-nonuniform distribution, most of the constants in \tabref{tab:breakdown} will change, but they will still be $\mathcal{O}(1)$ and will not change the overall communication complexity as shown in \tabref{tab:nonuniform}. The global tree and local tree can still be separated during the analysis for the $\mu$-nonuniform case. Teng \cite{Teng1998} proved that it is always possible to form a $P$-way partition of a $\mu$-nonuniform distribution with $\mathcal{O}(N/P)$ particles each. This means the depth of the global tree is always $\mathcal{O}(\log P)$ even for the $\mu$-nonuniform case. From Lemma 4.8 in Teng \cite{Teng1998} we know that the non-leaf-boxes have a in-degree bounded by a constant. Therefore, the number of cells per level for the global communication is $\mathcal{O}(1)$, and so are the number of processes to communicate with per level. This means that the communication complexity of both global M2L and global M2M phases is $\mathcal{O}(\log P)$.

In the uniform case, the local communications had a halo width of one and two cells for the P2P and M2L phases, respectively. In the general case, this depends on the multipole acceptance criteria (the definition of well-separatedness), but is bounded by a constant even for $\mu$-nonuniform distributions. Again, we may use Lemma 4.8 in Teng \cite{Teng1998} as a proof for this bound. Therefore, as long as the halo width is $\mathcal{O}(1)$ we can make the same surface to volume ratio argument to get a $\mathcal{O}(N/P)^{2/3}$ complexity as shown in \tabref{tab:nonuniform}. In conclusion, all the upper bounds on the communication complexity of the global M2L and M2M and the local M2L and P2P phases are valid for the $\mu$-nonuniform case. Therefore, the adaptive FMM with $\mu$-nonuniform distribution has a communication complexity of $\mathcal{O}(\log P+(N/P)^{2/3})$.

\section{Hierarchical Matrices as Algebraic Variants of FMM}
As mentioned in the introduction, hierarchical matrix representations of different flavors have been proposed in the last decade or so to exploit the underlying low rank hierarchical structure of matrices that appear in broad classes of applications. Hackbusch \cite{hackbusch99,hackbusch00} pioneered the concepts of hierarchical matrices in the form of $\mathcal{H}$ and $\mathcal{H}^2$ matrices and developed a substantial mathematical theory for their ability to approximate integral operators and boundary value problems of elliptic PDEs. There ideas have been developed considerably over the years, for the construction and use of hierarchical matrices in solving discretized integral equations and preconditioning finite element discretizations of PDEs~\cite{borm02, grasedyck03, borm05, leborne08, bebendorf09}. 
Hierarchically semi-separable (HSS) and hierarchically block-separable (HBR) are related and well-studied representations that also use low rank blocks of a dense matrix in a hierarchical fashion. The concept of a semi-separable matrix originated from matrices associated with separable kernels allowing their low rank representation as outer products of two thin matrices~\cite{golub05}. Matrices are semi-separable if their upper and lower triangular parts are, each, part of a low rank matrix. HSS matrices refer to matrices whose off-diagonal blocks are numerically of low rank and whose block structure consists of blocks that grow geometrically in size with their distance from the main diagonal. Fast factorization algorithms for HSS matrices have been developed in~\cite{chandra06,xia10}. HBR matrices have a similar structure but emphasize the telescoping nature of the matrix factorization \cite{gillman12} to use in the construction of direct solvers for integral equations \cite{martinsson13}.

A particular variant of hierarchical matrices, $\mathcal{H}^2$, has many similarities with Fast Multiple representations and can therefore benefit from the substantial algorithmic developments of FMM and in particular the $\mathcal{O}((N/P)^{2/3})$ communication complexity on distributed memory machines as derived in section 2.  In this section, we describe these similarities by showing how the matrix-vector multiplication operation in the hierarchical format maps to the various computational kernels of the FMM.

\subsection{Representation}
\label{rep}

Fig.~\ref{hrep} depicts the elements of an algebraic $\mathcal{H}^2$ representation of a dense matrix $A$. The representation consists of:   

\begin{figure}
\centering 
  \includegraphics[width=0.5\textwidth]{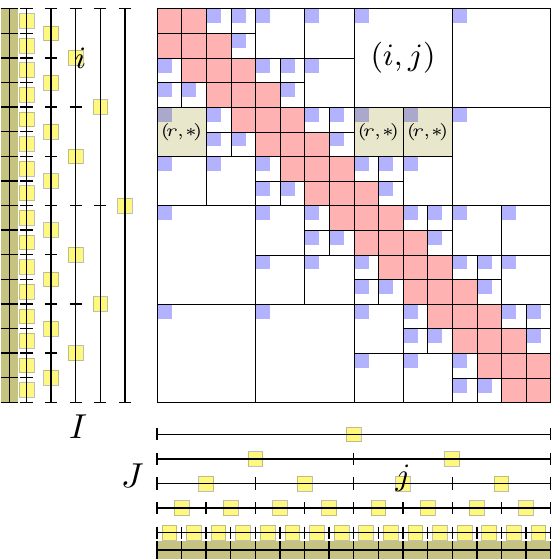}
  \caption{Hierarchical compressed representation of a dense matrix. Matrix blocks $A_{ij}$ are compressed by expressing them in factorized form $U_i S_{ij} V_j^t$.}
  \label{hrep}
\end{figure}

\begin{figure}
	\centering
  \includegraphics[width=\textwidth]{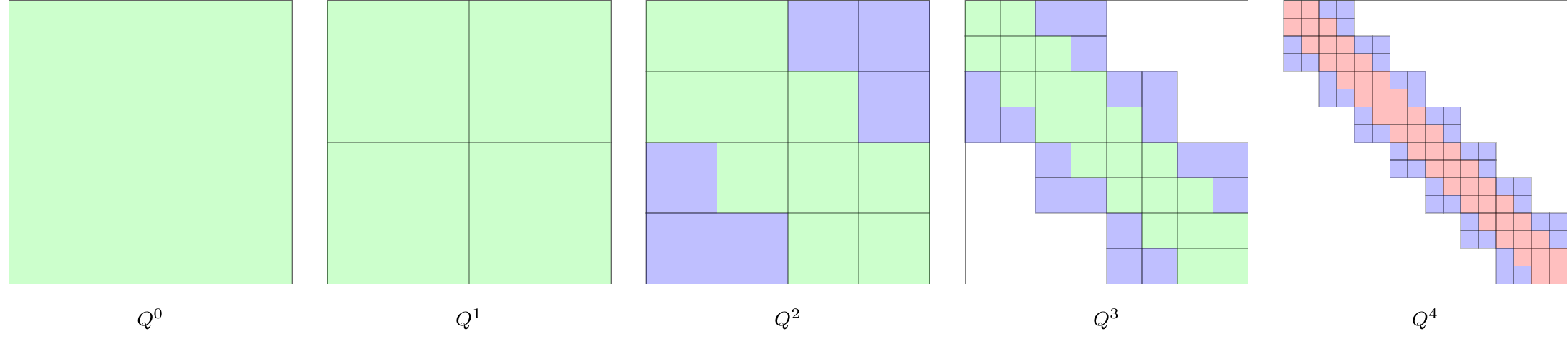}
  \caption{Quadtree representation of matrix structure. Leaves of the quadtree represent disjoint matrix blocks that collectively cover the complete matrix.}
  \label{qtree}
\end{figure}

\begin{itemize}
	\item a tree $I$ that organizes the row indices of the matrix hierarchically. We use a binary tree in the illustration in \figref{hrep} but a quadtree, or an octree, or another application-specific organization is possible. In particular, an octree is natural when the matrix is a discretization of a volume integral operator. The tree represents row blocks of the matrix. It is used to store column basis vectors in which to express the data of the various matrix blocks. The thin basis matrices $U$ are stored explicitly at its leaves, and small interlevel transfer matrices $E$ are stored at the higher levels and used to generate the level-appropriate bases as described below. 

	\item a tree $J$ that organizes the column indices of the matrix hierarchically. As with $I$, $J$ encodes possible block partitionings of the columns of $A$. It stores row basis vectors $V$ for the data of the matrix $A$. The basis matrices are stored at its leaves and interlevel transfer matrices $F$ are stored at the higher levels and used to generate dynamically the level-appropriate row basis.
		
	\item a quadtree $Q$ whose nodes are indexed by two indices $i \in I$ and $j \in J$. The $(i,j)$ pairs at the leaves of this incomplete quadtree (\figref{qtree}) form a hierarchical partitioning of the matrix. Collectively, the leaves cover the whole matrix $A$. Some of the leaves are labelled as low rank leaves and represent the blocks of $A$ that can be approximated to the desired tolerance by a low rank factorization $U_i S_{ij} V_j^t$. These leaves of the quadtree store the small matrices $S_{ij}$, which may interpreted as the projections of the $A_{ij}$ blocks on the corresponding $U_i$ and $V_j$ bases. 
	
	\item a set of dense $b\times b$ matrices that are not compressed. The complement of the low rank leaves of $Q$ represents blocks of $A$ that are not economically expressible as low rank factorizations and are more effectively stored in their original format. These leaves appear only at the lowest level of the quadtree $Q$ and represent blocks of sizes $b \times b$, where $b$ is tuned to size of the cache memories of the target hardware. In \figref{hrep} these dense blocks, for clarity of illustration, are shown along a diagonal band only but these blocks may appear anywhere in the matrix.   
\end{itemize}

\begin{figure}[h]
\centering 
  \includegraphics[width=0.3\textwidth]{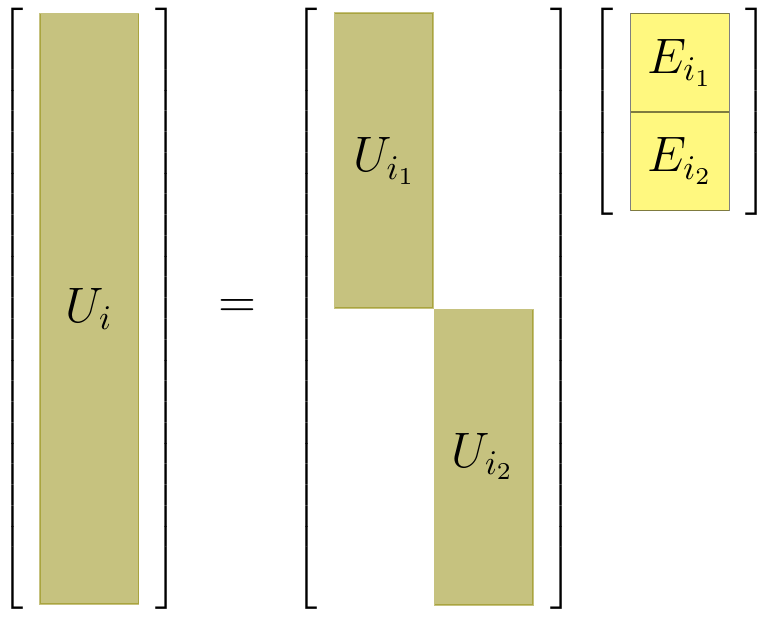}
  \caption{Hierarchical representation of column basis vectors of node $i$ in terms of its children. Low rank matrix blocks are expressed in terms of the level-appropriate basis vectors. A similar hierarchy exists for the row bases. }
  \label{hbasis}
\end{figure}

This representation of a matrix reduces the needed storage to $\mathcal{O}(N)$. This is due to the fact that all the blocks in a block row share a common column basis, and this basis is itself nested and can be expressed hierarchically. This is illustrated in \figref{hrep} where block row $r$ (shown as shaded) contains three blocks all sharing the same $U_r$ in their respective individual $U_r S_{rj} V_j^t$ representations. The index $r$ of this row block is not an index of a leaf node in the tree $I$ however; its column basis $U$ is not stored explicitly but is expressible in terms of the column bases of its children. Fig.~\ref{hbasis} shows the hierarchical structure of the column space. 

The growth of the storage requirements of this hierarchical representation can be estimated from the representations needed for storing the row and column bases and those needed for storing the leaves of the quadtree. Let $k$ be the rank used in the approximation of the $A_{ij} = U_i S_{ij} V_j^t$ blocks and assume for simplicity that it is constant across all blocks.  
The column bases requires $\mathcal{O}(N k)$ units of storage at the leaves and $\mathcal{O}(N k^2)$ for the transfer matrices 
The number of leaves of the quadtree is linear in $N$ assuming that the size of every block row at every level on the quadtree is bounded by some constant. Since every $S_{ij}$ matrix is of size $k\times k$, the storage of the quadtree requires $\mathcal{O}(N k^2)$ and therefore all the elements of the representation require storage that grows only linearly with matrix size.

\subsection{Matrix-Vector Multiplication with Hierarchical Matrices}

A matrix-vector operation with a hierarchical matrix can be performed in $O(N)$ arithmetic operations, and can use the same communication infrastructure of FMM. This may be seen by expressing the product as: 

\[
	y = \left( \sum_{(i,j) \in D} A_{ij} \right) x 
	    +   
	    \left( \sum_{(i,j) \in L}
	    U_i S_{ij}  V^t_j \right)  x  
	=  \underbrace{\sum_{(i,j) \in D} A_{ij} x_j}_{\substack{\text{Dense mat-vecs} \\ \text{operations}}} + 
		\underbrace{
		\sum_{i \in I} U_i
	    \underbrace{ 
			\sum_{(i,j) \in L}
			     S_{ij}  \underbrace{V^t_j \,  x}_{\text{Upsweep}}}_{\text{Coupling phase}} }_{\text{Downsweep}}
\]
where $D$ is the set of dense leaves of the quadtree and $L$ is the set of its low rank leaves. 

\begin{figure}[h]
\centering
	\includegraphics[width=3.5in]{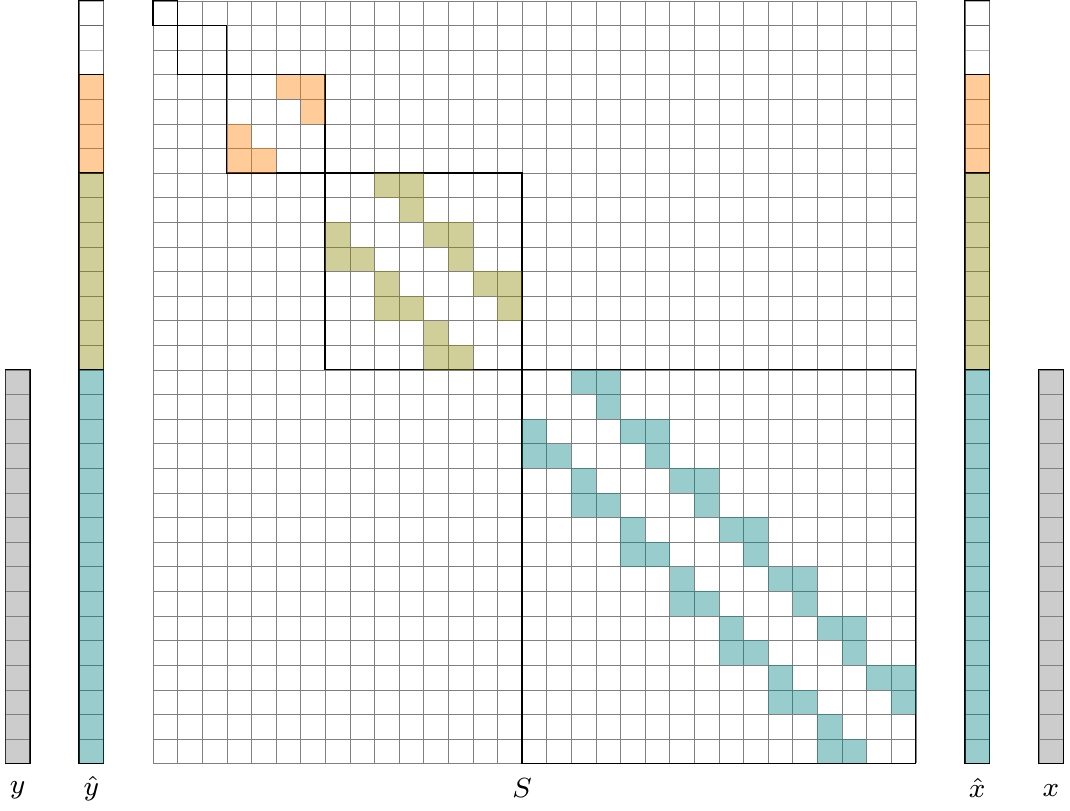}
	\caption{Low rank portion of the matrix-vector operation.}
    \label{matvec}
\end{figure}

The dense product portion consists of multiplications of $b\times b$ matrices with corresponding vectors, while the low rank portion of the operation can be separated into three computational kernels:  

\begin{description}
	
\item[Upsweep.] In this phase, the products $V_j^t x$  are computed for all $j \in J$. Since the bases $V_j$ are stored explicitly only at the leaves and the bases at interior nodes are expressible in terms of $k \times k$ transfer operators, this computation may be performed as follows: 
\begin{itemize}
	\item at the leaves: $\hat{x}_j = V_j^t x$
	\item at the interior nodes:  $\hat{x}_j = \sum F_k^t \hat{x}_k$
\end{itemize}
The result of the upsweep operations is a set of vectors of size $k \times 1$ at each node of the tree $J$ representing the matrix columns. This set of vectors is denoted by $\hat{x}$ and is shown (in a linearized fashion) in \figref{matvec}. 

\item[Coupling operations.] In this phase, the products $S_{ij}  \hat{x}_j$ are performed and accumulated as $\hat{y}_i$ in the corresponding nodes of the $I$ tree. This is illustrated in \figref{matvec} where the three colors are used to distinguish the $S_{ij}$ corresponding to the three levels $Q^2$, $Q^3$, and $Q^4$ of the matrix in \figref{qtree}. This phase may be viewed as a set of block-sparse matrix-vector multiplications, one per level of the tree. The various levels of may be processed concurrently, and the block sizes in these block-sparse matrices are $k\times k$. 

\item[Downsweep.] In this final portion of the computation the products $U_i \hat{y}_i$ are performed and assembled in their corresponding positions in the final vector $y$. Since the bases $U_i$ are available in explicit form only at the leaves, all the products that correspond to interior nodes have to be computed by going down the $I$ tree. Starting the root, contributions to the children of every interior node are  accumulated and a direct computation is performed at the leaves. The computational pattern is dual to that of the upsweep phase: 
\begin{itemize}
	\item at interior nodes: $\hat{y}_k \ += E_k \hat{y}_i$  for every child $k$ of node $i$
	\item at leaf nodes: $y = U_i \hat{y}_i$ 
\end{itemize}
\end{description} 

\begin{table}[b]
\caption{Communication breakdown of hierarchical matrix-vector multiplication compared to FMM (cf. \tabref{tab:breakdown}).}
\label{tab:matvec}
\centering
\begin{tabular}{|l|l|c|c|c|c|}
\hline
\shortstack{H-matrix \\ operation}  & \shortstack{FMM \\ operation} & Processes & Blocks per level & Blocks per Process & Communication\\
\hline \hline
\shortstack{Global \\ $\sum S_{ij} \hat{x}_j$} & \shortstack{Global \\ M2L} & $\displaystyle\sum_i^{\log P}\mathcal{O}(1)$ & $\mathcal{O}(1)$ & $\mathcal{O}(1)$ & $\mathcal{O}(\log P)$\\
\hline
\shortstack{Global \\ $\sum  F_k^t \hat{x}_k$} & \shortstack{Global \\ M2M} & $\displaystyle\sum_i^{\log P}\mathcal{O}(1)$ & $\mathcal{O}(1)$ & $\mathcal{O}(1)$ & $\mathcal{O}(\log P)$\\
\hline
\shortstack{Local \\ $\sum S_{ij} \hat{x}_j$} & \shortstack{Local \\ M2L} & $\mathcal{O}(1)$ & $\mathcal{O}(2^{(d-1) i})$ & $\displaystyle\sum_i^{\log_{2^d}(N/P)}\mathcal{O}(2^{(d-1) i})$ & $\mathcal{O}((N/P)^{\frac{d-1}{d}})$\\
\hline
\shortstack{Local \\ $\sum A_{ij} x_j$} & \shortstack{Local \\ P2P} & $\mathcal{O}(1)$ & $\mathcal{O}(2^{(d-1) i})$ & $\mathcal{O}(2^{(d-1) \log_{2^d}(N/P) }) $ & $\mathcal{O}((N/P)^{\frac{d-1}{d}})$\\
\hline
\end{tabular}
\end{table}

The analogy to the FMM computations and communication patterns is partially summarized in \tabref{tab:matvec}. The upsweep phase in the hierarchical matrix-vector multiplication corresponds to building and propagating multipole expansions (P2M and M2M) up the tree. The block multiplications of the coupling phase correspond to computing M2L interactions. The downsweep phase corresponds to propagating local expansions (L2L and L2P). Finally the dense $b\times b$ multiplications correspond to P2P direct interactions between nearfield particles. 

The trees $I$ and $J$ may be assumed to have a constant ($2^d$) number of children per node resulting in binary trees, quadtrees, or octrees depending on the dimension $d$. The most natural type of tree to use depends on the provenance of the matrix. For problems involving the discretization of volume integrals, octrees are perhaps the most reasonable ones to use. Schur complement matrices arising from planar interfaces may be more naturally represented using quadtrees even in volumetric problems. Evidently, it is always possible to use binary trees regardless of the origin of the matrix, but this choice may limit attainable performance. The $I$ and $J$ trees may not be uniform but may have an adaptive depth as warranted by the structure of the matrix. 

The structure of $Q$, the matrix quadtree in which the $S$ blocks are stored as illustrated in \figref{qtree}, does not generally have the same regularity as the corresponding set of interaction lists of FMM computations. While FMM methods use a strictly-geometric criterion to determine admissibility, hierarchical matrices may have a general structure for the sparsity of these block rows reflecting a non-geometric admissibility criterion. This is in fact the specific sense in which this hierarchical representation can be viewed as an algebraic variant of FMM. The leaves of the $Q$ tree are generated algebraically so that a factorization  $U_i S_{ij} V_j^t$ of rank-$k$ results in an appropriate approximation of the block $A_{ij}$. FMM computations generally do not store the $S_{ij}$ blocks but rather compute them from their analytical expressions at run-time as needed for the M2L interactions, hence saving $\mathcal{O}(Nk^2)$ storage. Such savings are only possible when there is an underlying analytical kernel. For the more general problems, storing the $k\times k$ $S_{ij}$ matrices at the leaves of the $Q$ quadtree is what allows hierarchical matrix representations to approximate general (dense) operators in a way that is scalable both arithmetically and in terms of communication.  

In order to obtain meaningful bounds on the communication complexity of the hierarchical matrix vector multiplication, we assume that the number of blocks in a block row is bounded by a constant independent of $N$. In FMM computations using a regular Cartesian subdivision, this number corresponds to the number of same-size cells that are interacted with and hence is the same constant for all levels in that case. Having an $\mathcal{O}(1)$ size for the block rows is essential for maintaining linear arithmetic complexity.  We also assume that the bandwidth of any level in the quadtree $Q$ is bounded by a constant. If we denote by $d$ the spatial dimension of the problem that generated the matrix, the amount of communication may require a large halo, due to a potentially large bandwidth in some row-block, but of $\mathcal{O}(1)$. The asymptotic complexity therefore still scales as the surface to volume ratio of $d$-dimensional blocks, using similar arguments to those described in Section 2. The first and third rows of \tabref{tab:matvec} show the asymptotic communication results for the global and local portions of the coupling phase for problems originating from a (small) spatial dimension $d$. The global and local portions of this computation are distributed as described in \figref{fig:communication} in section 2. 

The upsweep and downsweep phases of the multiplication do not depend on the particular structure of the $Q$ quadtree but only on the structure of the $I$ and $J$ trees. For practical implementations, these trees have a regular fanout of $2^d$, and although they may be nonuniform in their depth they do not depend on the bandwidth of $Q$. Their asymptotic communication complexity is therefore essentially the same as that of the FMM for both the uniform and non-uniform cases as described in section 2. The second row of \tabref{tab:matvec} displays identical results to \tabref{tab:nonuniform} for the global portion of the upsweep phase. The asymptotic constants however would depend on $d$ and on the nonuniformity of the tree depth. The rest of the upsweep and downsweep phases have analogous results. Finally, the matrix blocks requiring dense $b\times b$ direct block multiplications may induce more communication than a regular FMM P2P phase due the distribution of the dense  blocks at the lowest level, but the bounded bandwidth insures a communication cost similar to that of the coupling phase. In short, this hierarchical representation under fairly reasonable assumption on its structure inherits the strong communication complexity results that have made FMM a powerful kernel for extreme computing.

\section{Conclusions}
Driven by seemingly inexorable trends in computer architecture at extreme scale, we have identified an algebraic form of the fast multipole method (AFM) as a candidate for migrating basic linear algebraic subroutines to hybrid hierarchical distributed-shared memory machines targeting billion-thread concurrency, where the performance portability of trusted workhorse methods of bulk synchronous structure is questionable, due to the lack of performance guarantees of individual cores. AFM is a hybrid of the fast multipole and $\mathcal{H}$-matrices, with the algebraic generality of the latter captured in the distributed tree-like data structures of the former. As with any method predicated upon compression due to low rank, the mathematical efficiency of AFM depends upon the operator. AFM is of interest for both dense and sparse operators arising from physical models through the discretization of integral and differential equations, where the underlying solution fields are fundamentally continuous and the operators smoothly varying functions of distance, with sufficiently rapid decay.  AFM does not presume possession of a Green's function. With respect to massively distributed memory, it benefits from the FMM feature that no all-to-all synchronization is required. With respect to accelerators such as GPGPUs that impose in the hardware their own scales, such as the number of threads in a warp that must execute in a SIMT manner, it benefits from the tradeoff between the order of the expansion and the separation of the interacting degrees of freedom. One may choose an order that fits the architecture and enforce the admissibility condition for low-rank representation of an interaction by the size of the blocks.  

The primary claims of this contribution are:
\begin{itemize}
\item The communication complexity of FMM for uniform distributions is $\mathcal{O}\left(\left(\frac{n}{p}\right)^\alpha+\log P\right)$, where $\alpha=2/3$ for three dimensions.
\item This complexity still holds for nonuniform distributions with adaptive tree structures to which FMM is applicable, and is generalizable to arbitrary dimensions.
\item This complexity also holds for algebraic variants of FMM, such as H-matrices, HSS, and RS, wherever an upper bound exists on the number of blocks in a block row in the low-rank representation.
\end{itemize}

We have not yet built an implementation of the algebraic fast multipole method that illustrates of its potential performance advantages and versatility.  There are components that need to be developed at the GPGPU or accelerator scale and the resulting node code needs to be merged with the FMM tree data structure and traversal mechanisms.  The memory efficiency and strong absolute and scaling performance of FMM for analytically specified kernels such as the Laplacian is well documented in existing implementations, such as ExaFMM, so the primary criterion affecting the memory consumption and execution time of the AFM as a solver for linear systems or as a component, such as a matrix-vector multiply or a preconditioner in a larger context, is how efficiently general operators of interest can be compressed. We conclude with four ripe open problems:
\begin{itemize}
\item
For operators that do not satisfy the requirement of bounded block bandwidth, independent of $N$, in their low-rank representation, how does the communication complexity of AFM generalize beyond the convenient bound inherited from FMM?  What is the analog of dimension $d$ in this case?  
\item 
What are the sizes of constants in the asymptotic communication complexity of the FMM and AFM method? 
These will generally depend upon the distributions of interacting degrees of freedom represented in the tree data structures and mathematical ranks of the operators involved.  These constants will determine the natural crossover points for the applicability of FMM and AFM relative to other methods that excel in smaller problems but possess inferior asymptotic communication complexity.
\item
What is the achievable asynchronicity of the communications of FMM and AFM in practice and in what programming models are they best expressed?
\item
How should the algorithmic hierarchy be adapted to architectural hierarchy in practice?  In general, the ``natural'' granularities (e.g., FMM expansion degree and number of threads in a GPGPU warp, size of AFM bases and capacities of various levels of cache) will not match, and the mismatch has performance implications.  This highlights the importance of the tunability of the granularities of hierarchy that are at the disposal of the algorithm designer and scientific user and their exposure in the software.
\end{itemize}

\section*{Acknowledgments}
Our inspiration for this paper comes from the community of researchers who joined us for two workshops bearing the title ``Scalable Hierarchical Algorithms for eXtreme Computing'' (SHAXC), held at the King Abdullah University of Science and Technology (KAUST) during 28-30 April 2012 and 2-4 May 2014.  We are especially grateful to Lorena Barba and Matthew Knepley, who joined us in co-organizing SHAXC-1, and to Alex Litvenenko, who joined us in co-organizing SHAXC-2.  Sponsorship of SHAXC-1 was from the Strategic Initiative in Extreme Computing and of SHAXC-2 from its successor, the Extreme Computing Research Center (ECRC) at KAUST.  Archives of the workshops may be found online. One of ten research centers at KAUST, the ECRC operates on two fronts: (1) performing basic research and developing algorithms and software that will enable today's applications to migrate to the exascale systems that are expected by the end of the decade, and (2) enabling KAUST's scientific and engineering simulation campaigns to exploit today's state-of-the-practice petascale systems.

\bibliographystyle{plain}

\end{document}